\documentclass{article}

 \usepackage[final,nonatbib]{bdl_2018}

\usepackage[utf8]{inputenc} % allow utf-8 input
\usepackage[T1]{fontenc}    % use 8-bit T1 fonts
\usepackage[hidelinks]{hyperref}       % hyperlinks
\usepackage{url}            % simple URL typesetting
\usepackage{booktabs}       % professional-quality tables
\usepackage{amsfonts}       % blackboard math symbols
\usepackage{nicefrac}       % compact symbols for 1/2, etc.
\usepackage{microtype}      % microtypography
\usepackage{graphicx}
\usepackage[numbers]{natbib}
\usepackage{mathtools}
\usepackage{amsmath}
\usepackage{amssymb}
\usepackage{courier}
\usepackage{siunitx}
\usepackage{xcolor}
\usepackage{bm}
\usepackage{float}
\usepackage{caption} 
\title{Bayesian Deep Learning for\\ Exoplanet Atmospheric Retrieval}

\author{Frank Soboczenski,$^{a}$ Michael D. Himes,$^{b}$ Molly D. O'Beirne,$^{c}$ Simone Zorzan$^{d}$\vspace*{0mm}\\
\textbf{Atılım Güneş Baydin,$^{e}$ Adam D. Cobb,$^{e}$ Yarin Gal,$^{f}$ Daniel Angerhausen$^{g}$}\vspace*{0mm}\\
\textbf{Massimo Mascaro,$^{h}$ Giada N. Arney,$^{i,j}$ Shawn D. Domagal-Goldman$^{i,k}$}\vspace*{0mm}\\
$^{a}$School of Population Health and Environmental Sciences, King's College London\\
$^{b}$Planetary Sciences Group, Department of Physics, University of Central Florida\\
$^{c}$Department of Geology and Environmental Science, University of Pittsburgh\\
$^{d}$ERIN Department, Luxembourg Institute of Science and Technology\\
$^{e}$Department of Engineering Science, University of Oxford\\
$^{f}$Department of Computer Science, University of Oxford\\
$^{g}$Center  for  Space  and  Habitability,  University  of  Bern\\
$^{h}$OCTO Applied AI, Google Cloud\\
$^{i}$Virtual Planetary Laboratory Team, NASA Astrobiology Institute\\
$^{j}$Planetary Systems Laboratory, NASA Goddard Space Flight Center\\
$^{k}$Planetary Environments Laboratory, NASA Goddard Space Flight Center\\
\texttt{frank.soboczenski@kcl.ac.uk; mhimes@knights.ucf.edu}\\
\texttt{mdobeirne@pitt.edu; simone.zorzan@list.lu}\\
\texttt{\{gunes, acobb\}@robots.ox.ac.uk; yarin@cs.ox.ac.uk}\\
\texttt{daniel.angerhausen@csh.unibe.ch; massy@google.com}\\
\texttt{\{giada.n.arney, shawn.goldman\}@nasa.gov}
}
\begin{document}
% \nipsfinalcopy is no longer used

\maketitle

%\begin{abstract}
%  The abstract paragraph should be indented \nicefrac{1}{2}~inch
%  (3~picas) on both the left- and right-hand margins. Use 10~point
%  type, with a vertical spacing (leading) of 11~points.  The word
%  \textbf{Abstract} must be centered, bold, and in point size 12. Two
%  line spaces precede the abstract. The abstract must be limited to
%  one paragraph.
%\end{abstract}

\section{Introduction}

Over the past decade, the study of extrasolar planets has evolved rapidly from plain detection and identification to comprehensive categorization and characterization of exoplanet systems and their atmospheres. Atmospheric retrieval, the inverse modeling technique used to determine an exoplanetary atmosphere’s temperature structure and composition from an observed spectrum, is both time-consuming and compute-intensive, requiring complex algorithms that compare thousands to millions of atmospheric models to the observational data to find the most probable values and associated uncertainties for each model parameter \citep{madhusudhan2018}. For rocky, terrestrial planets, the retrieved atmospheric composition can give insight into the surface fluxes of gaseous species necessary to maintain the stability of that atmosphere, which may in turn provide insight into the geological and/or biological processes active on the planet \citep{SchwietermanEtal2017asbioBiosignaturesReview}. These atmospheres contain many molecules, some of them biosignatures, spectral fingerprints indicative of biological activity, which will become observable with the next generation of telescopes \citep{FujiiEtal2018asbioBiosignatures}. Runtimes of traditional retrieval models scale with the number of model parameters, so as more molecular species are considered, runtimes can become prohibitively long. Recent advances in machine learning (ML) and computer vision \citep{krizhevsky2012,he2016deep} offer new ways to reduce the time to perform a retrieval by orders of magnitude \citep{marquez2018supervised, zingales2018}, given a sufficient data set to train with. Here we present an ML-based retrieval framework called \emph{Intelligent exoplaNet Atmospheric RetrievAl} (INARA) that consists of a Bayesian deep learning model for retrieval and a data set of 3,000,000 synthetic rocky exoplanetary spectra generated using the NASA Planetary Spectrum Generator (PSG) \citep{villanueva2018}. Our work represents the first ML retrieval model for rocky, terrestrial exoplanets and the first synthetic data set of terrestrial spectra generated at this scale.

\section{Background}
\label{gen_inst}

Traditionally, the study of exoplanetary atmospheres has been done by fitting forward models to observational data, which is based on the relative decrease in flux when the exoplanet is in front of or behind its host star \citep{Crossfield2015paspAtmospheres, DemingSeager2017jgreExoplanetAtmospheres}. This is usually performed using a Monte Carlo sampling method in a Bayesian framework to propose atmospheric models, simulate the spectrum, and compare it to the observed data \citep[e.g.,][]{skilling2004, terBraak2008}. Degeneracies among atmospheric parameters complicate this process, necessitating the evaluation of hundreds of thousands to millions of atmospheric models to fully explore the parameter space. This results in a posterior distribution which characterizes these degeneracies and informs the relative probability of the ranges of values considered for each model parameter. While these sampling methods are executed in parallel, this task still requires a significant amount of computational time \citep{madhusudhan2018}.

Recently, the exoplanet community has begun to apply supervised ML methods to the problem of atmospheric retrieval. Waldmann \citep{Waldmann2016apjDreamingAtmospheres} used a deep belief network to identify molecular species in an observed spectrum, paving the way for more advanced ML applications. Building upon this, two ML retrieval algorithms have been developed to date: ExoGAN \citep{zingales2018} and HELA \citep{marquez2018supervised}. These produce results in seconds to minutes, compared to on the order of 100 CPU hours for the aforementioned traditional Monte Carlo sampling-based methods. ExoGAN utilizes a generative adversarial network (GAN) \citep{goodfellow2014generative} to approximate the data distribution of realistic spectra and then uses the trained GAN to make predictions using inpainting to infer planetary conditions from observed spectra. HELA uses random forests \citep{ho1995random} to similarly make predictions of planetary parameters from observed spectra. Both models produce results that are generally consistent with conventional retrieval methods. Note that these models specialize in hot Jupiters, a class of gas giant exoplanets on very short-period orbits, and consider less than a handful of molecules.

\begin{figure}
  \centering
  \includegraphics[width=\textwidth]{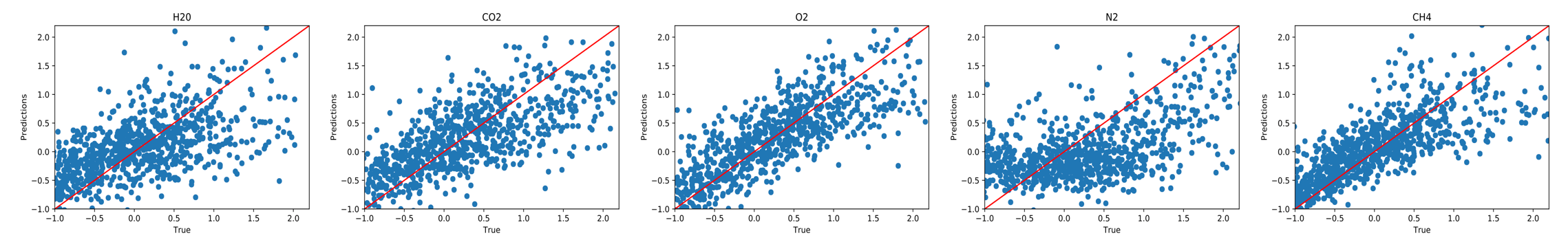}\hfill
  \includegraphics[width=0.33\textwidth]{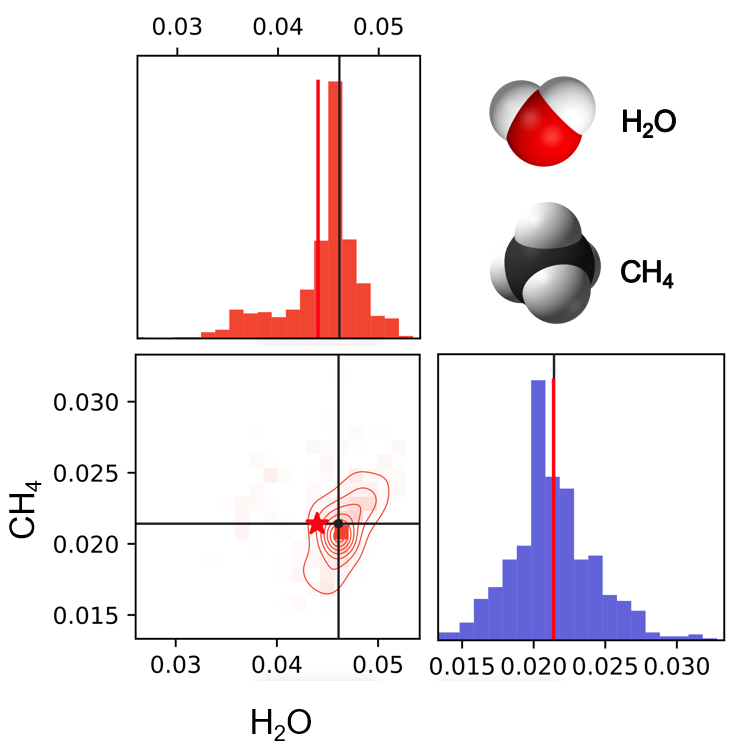}\qquad\qquad\qquad
  \includegraphics[width=0.33\textwidth]{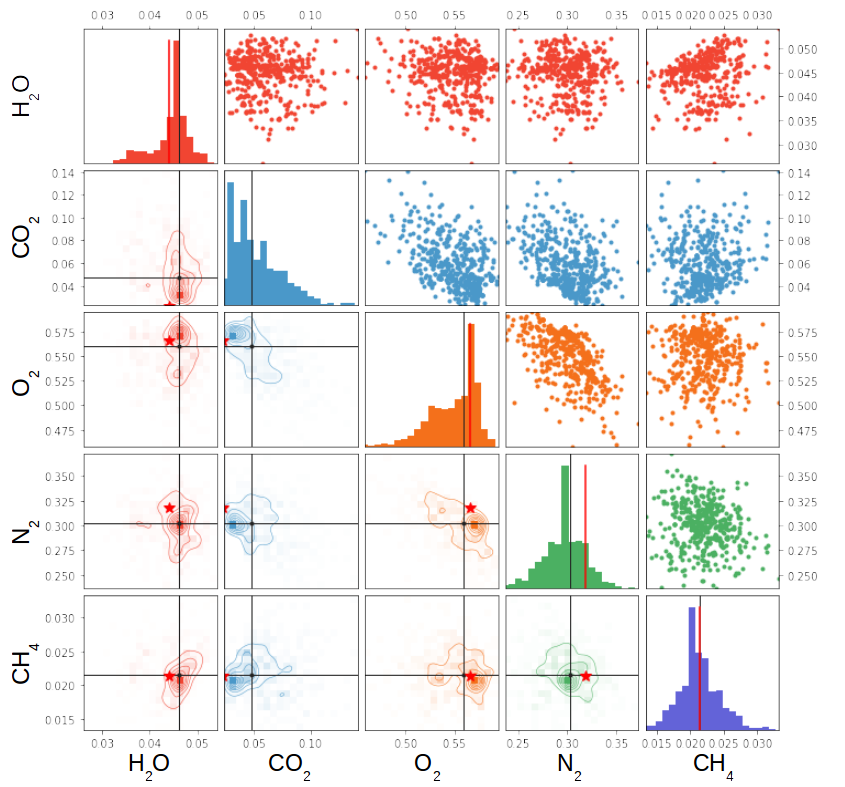}
  \caption{Predictions of H$_2$O, CO$_2$, O$_2$, N$_2$, CH$_4$ based on the best performing model; training limited to 64 epochs on 110,000 parameter–spectra pairs hence some uncertainties reflect calibration issues.}
  \label{fig:results}
\end{figure}

\section{Methods}

We train a deep neural network in a supervised setting to predict exoplanet atmospheric parameters $\bm{\theta}$ given an observed spectrum $\bm{s}$, using a training set $(\bm{s}_i, \bm{\theta}_i), i=1,\dots,m$ that we generate by running the NASA PSG\footnote{\url{https://psg.gsfc.nasa.gov/}} \citep{villanueva2018} simulator to get $\bm{s}_i = \mathrm{psg}(\bm{\theta}_i)$, where the parameters are sampled from a physically-motivated prior model $\bm{\theta}_i \sim p(\bm{\theta})$. Spectra $\bm{s}_i$ are vectors (of length 4379) describing radiation intensity as a scalar function of wavelength and therefore we explore a series of 1D convolutional neural network (CNN) configurations. In order to train our CNN models, we generate a data set encompassing spectra based on a given planetary system model, where we consider F-, G-, K-, and M-type main sequence stars. Observations are simulated using an instrument model of the Large UltraViolet/Optical/InfraRed Surveyor (LUVOIR), a design concept for a multi-wavelength space observatory, but with a much higher resolution. The prior model $p(\bm{\theta})$ comprises planetary parameters (radius, mass, surface pressure, semi-major axis, pressure temperature profile) and atmospheric compositions. Planetary parameters are randomly selected from ranges and distributions consistent with our solar system and observations of other systems \citep{boyajian2013, boyajian2012, robinson2014, rogers2015, sotin2007, Zahnle2017}.  The ranges for these parameters are chosen such that a planet in an Earth-like orbit can vary in temperature by a few hundred Kelvin. We consider 12 molecules based on the composition of atmospheres in our solar system as well as the observability of species \citep{FujiiEtal2018asbioBiosignatures}:  H$_2$O, CO$_2$, O$_2$, N$_2$, CH$_4$, N$_2$O, CO, O$_3$, SO$_2$, NH$_3$, C$_2$H$_6$, and NO$_2$. Concentrations are randomly selected within a range based on the observed composition of atmospheres in our solar system. While cloud mixing ratios are calculated, clouds are ignored in our simulations due to the computational burden as even poor modeling efforts increase computational time by a factor of 50.  

We use the Monte Carlo dropout approximation to produce predictive distributions over parameters $\bm{\theta}$. Dropout is a common regularization technique in neural networks to prevent overfitting and allow for a more generalizable model \citep{hinton2012}. It has recently been shown that applying dropout at both training and test time is equivalent to making a variational approximation to the posterior distribution over the network weights \citep{gal2016}. Each dropout mask removes a certain proportion of weights by setting them to zero during a forward pass. Therefore, multiple forward passes with different dropout masks for the same input gives a set of predictive samples that build a predictive distribution. Through implementing dropout both at training and test time, we are effectively sampling from the posterior over weights of the network. This distribution over the weights enables us to approximate a predictive distribution $p(\bm{\theta}\vert\bm{s})$ over the parameters of an exoplanet given an observed spectrum.

INARA is implemented in Python using PyTorch, and the source code and a Docker image are publicly available.\footnote{\url{https://gitlab.com/frontierdevelopmentlab/astrobiology/inara}} To interact with NASA Goddard PSG, the simulator at the core of our spectrum generation setup, we implemented a Python package called \emph{pypsg}\footnote{\url{https://gitlab.com/frontierdevelopmentlab/astrobiology/pypsg}} that handles data generation in PSG format and http-based two-way communication with PSG servers. The INARA codebase covers the running of server instances for data generation, ML model training, and inference in a distributed fashion utilizing the Google Cloud infrastructure. For the generation of the data set of 3M parameter and spectrum pairs, we employed approximately 2,000 high-end VMs (groups of 16 INARA instances connected to one PSG node). %Figure \ref{fig:overview} gives an overview of the distributed architecture.

\section{Preliminary Results, Discussion, and New Horizons}

\begin{table}
\centering
\caption{Comparison of atmospheric retrieval methods.}
\begin{tabular}{lll}
\toprule
Method & CPU time for inference & Number of molecules retrieved\\
\midrule
Traditional & Hundreds of hours & User-specified\\
ExoGAN \citep{zingales2018} & Minutes & 4\\
HELA \citep{marquez2018supervised} & Seconds & 3\\
INARA & Seconds & 12\\
\bottomrule
\end{tabular}
\label{tbl:inara-comparison}
\end{table}

We performed a grid search over model architectures and training hyperparameters, exploring over 70 combinations of different architectures (linear regression, feed-forward neural networks, and CNNs), learning rates in $\left[0.0001, 0.01\right]$, activation functions in \{tanh, ReLU, ELU\}, and optimization algorithms \{ADAM, SGD, ADAdelta, RMSProp\}. No dropout was used in this phase. Due to time constraints, model training was set to 64 epochs in all cases, using a training set of 110,000 parameter--spectra pairs. We used a mean square error (MSE) loss, and employed early stopping with a validation set of size 10,000 to avoid overfitting. 1D CNNs produced the best results, and we settled on a model with the configuration Conv1d(64)--tanh--MaxPool--Conv1d(64)--relu--MaxPool--Conv1d(128)--relu--MaxPool--Conv1d(256)--relu--FC(256)--relu--FC(12), which has approximately 18M trainable parameters.

Here we report results of the best 1D CNN model trained using 110,000 parameter--spectra pairs out of the 3M data set, leaving results with the full training set to future work. The prediction for 1,000 spectra is illustrated in Figure~\ref{fig:results}. H$_2$O, CO$_2$, O$_2$, N$_2$ and CH$_4$ are shown in the five plots in Figure~\ref{fig:results} top row, where each dot represents the average of 600 runs of our model with dropout for each planet. The details for predictive joint distributions for a random planet among those simulated is shown in the two bottom plots of Figure~\ref{fig:results}. The true value, indicated by the red star and the red line, falls within the predictive distribution for both parameters. Figure~\ref{fig:results_extended} in the appendix presents predictions for the full set of 12 molecules. INARA outperforms traditional Monte Carlo-based approaches by several orders of magnitude while computing a larger set of parameters and atmospheric molecules (Table \ref{tbl:inara-comparison}).

Thanks to the computational resources we had access to, our present data set is the largest collection of rocky planet spectra to date. For the first time in ML atmospheric retrieval, we adopted Monte Carlo dropout \citep{gal2016}, providing a predictive distribution comparable to the posterior distributions yielded by traditional, Bayesian approaches. Further investigation is necessary to determine how this predictive distribution compares to the posterior distributions of traditional methods. While we obtained good results, our search for the best model is incomplete, and a thorough exploration of different neural network architectures is desirable. In addition, a more detailed data set (i.e., in terms of wavelength, self-consistency, and the presence of clouds/hazes) could be used with INARA to generate more reliable and scientifically-informative models.\\

\section*{Acknowledgements}

This work originated at the NASA Frontier Development Lab (FDL), an accelerated research program focused on finding solutions for space-related scientific challenges using ML, with support by NASA, the SETI Institute, and industry partners Google Cloud, IBM, Intel, Nvidia, XPRIZE, KBRwyle, KX, Luxembourg Space Resources, and Lockheed Martin. We thank the NASA Astrobiology Institute for their support of the astrobiology challenges at FDL 2018; our amazing mentors for their fantastic support and guidance; Geronimo Villanueva at NASA Goddard Space Flight Center for offering extensive support in setting up PSG on Google Cloud; Sara Jennings, Shyla Spicer, and James Parr for organizing NASA FDL; the SETI Institute and NASA Ames Research Center for hosting us; and all industry partners involved in the program. 

% \section*{References}
% \medskip
\bibliographystyle{abbrv}
{\small
\bibliography{main}}

\newpage
\section*{Appendix}
\begin{figure}[h]
  \centering
  \includegraphics[width=\textwidth]{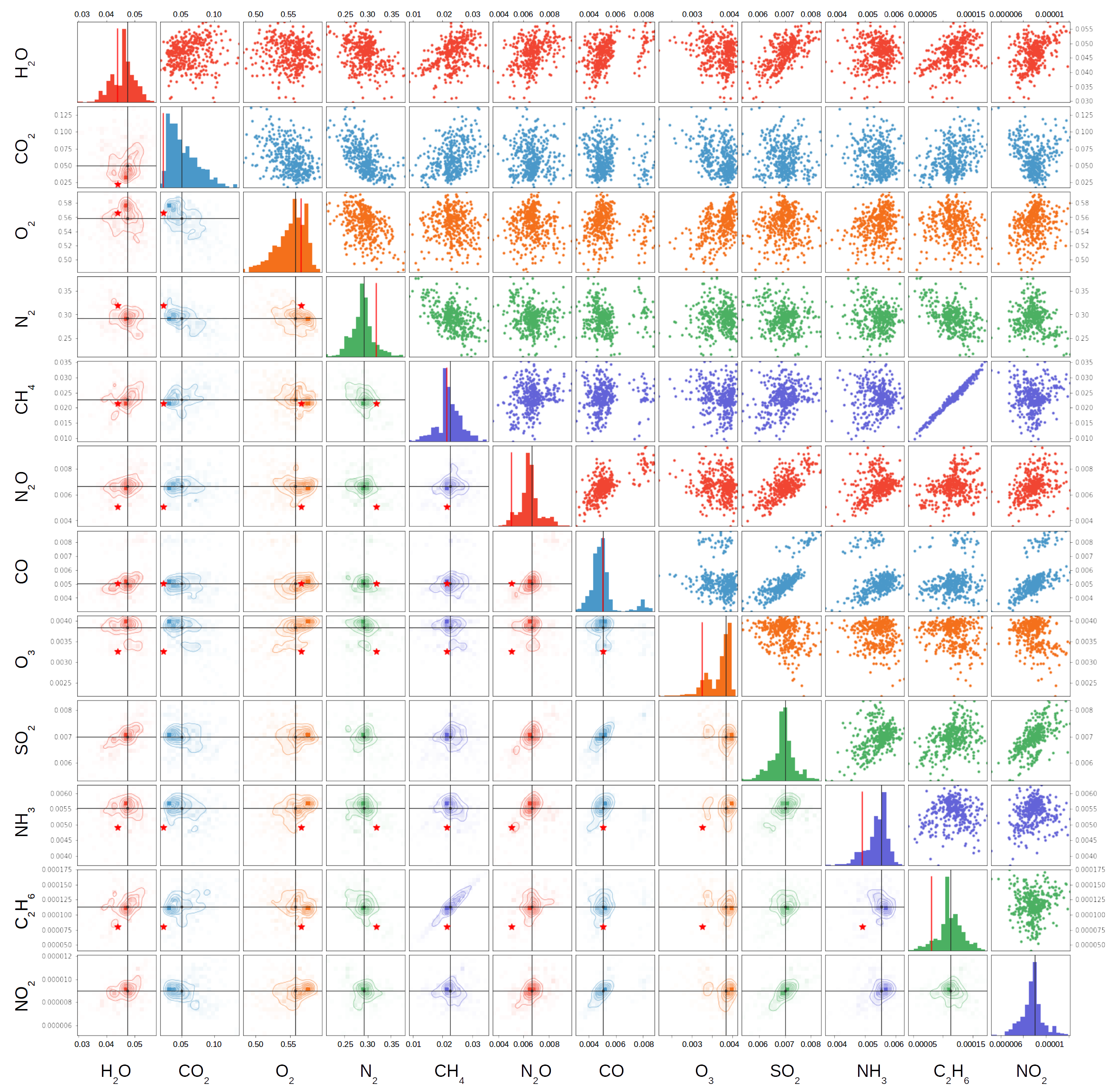}
  \caption{Posterior results with all 12 molecules in the model.}
  \label{fig:results_extended}  
\end{figure}

\end{document}